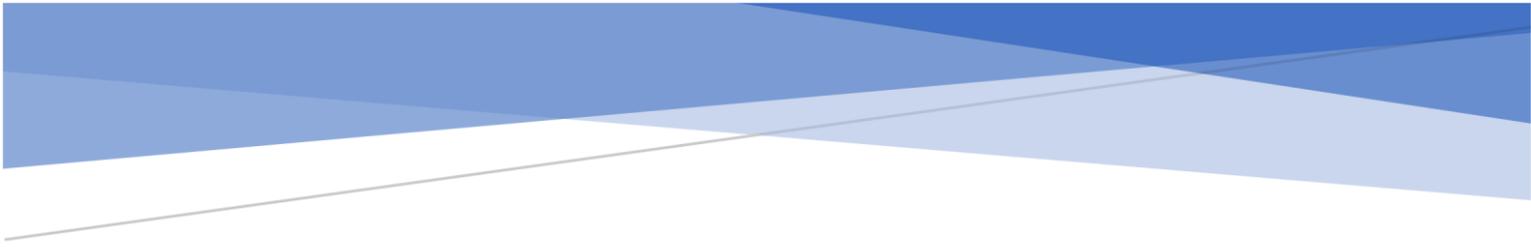

# 6G WHITE PAPER: RESEARCH CHALLENGES FOR TRUST, SECURITY AND PRIVACY

Document history:
*v. 0.1. Feb. 1$^{st}$ 2020*: First combined document draft structure with placeholders for chapter contributions.
*v. 0.2. Feb. 12$^{th}$ 2020*: Modifications to structure based on telco 7$^{th}$ Feb and initial contributions merged into one single document.
*v. 0.3. Mar. 2$^{nd}$ 2020*: Section edits merged into one single document and executive summary added. Document ready for first internal review.
*v. 0.4. Mar. 8$^{th}$ 2020*: Document ready for external review of 6G Summit participants
*v. 0.5. Apr. 7$^{th}$ 2020*: Revision after 6G summit
v. 1.0. Apr. 24$^{nd}$ 2020: Document draft ready to be submitted to arXiv

*Editor in chief: Mika Ylianttila*
*Section editors: Raimo Kantola, Andrei Gurtov, Lozenzo Mucchi, Ian Oppermann*

# Table of Contents

## Executive summary: main research challenges

**Vision: Trustworthy 6G.** The challenges in creating a trustworthy 6G are multidisciplinary spanning technology, regulation, techno-economics, politics and ethics. A combination of the current regulation, economic incentives and technology are maintaining the current level of hacking, lack of trust, privacy and security on the Internet. In 6G, this will not suffice, because physical safety will more and more depend on information technology and the networks we use for communication. Therefore, we need trustworthy 6G. The roles of trust, security and privacy are somewhat interconnected, but different facets of next generation networks. This white paper addresses their fundamental research challenges.

**Research challenge 1: Inherited and novel threats in 6G scale.** The diversity and volume of novel IoT devices and their control systems will continue to pose significant security and privacy risks and additional threat vectors as we move from 5G to beyond towards 6G system. The volume of new IoT devices introduced into 6G network will increase 10x from 10 billion scale of 5G networks to 100 billion scale in 6G. As a result of such deployment and use of 6G, the dependence of the economy and societies on IT and the networks will deepen. Safety will depend on IT and the networks. The development of AI blurs the line between reality and fake content and helps to create ever more intelligent attacks. The role of IT and the networks in national security keeps rising – a continuation of what we see in 5G.

**Research challenge 2: End-to-end trust in 6G.** In current "open internet" regulation, the telco cloud can be used for trust services only equally for all users. 6G should position the future cellular network as a solution to the all issues of trustworthy or trust networking such that network based information technology can be trusted to provide expected outcomes even in the face of malicious actors trying to interfere. 6G network must support embedded trust such that the resulting level of information security in 6G and the packet data networks where 6G provides connectivity to is significantly better than in state-of-the art networks commonly used today. Trust modeling, trust policies and trust mechanisms need to be defined.

**Research challenge 3: Post-quantum cryptography and security architecture for 6G.** The current 5G standard does not address the issue of quantum computing but relies on traditional cryptography. The development towards cloud and edge native infrastructures is expected to continue in 6G networks. While large-scale quantum computing can be expected to take longer, it is time to prepare for the shift to cryptography that is secure in the post-quantum world. According to current knowledge, contemporary symmetric cryptography remains secure for the most part even after the advent of quantum computing. Future of SIM cards and use of asymmetric cryptography will be interesting research questions.

**Research challenge 4: Machine-learning as tool and risk in softwarized 6G.** As 6G moves toward THz spectrum with much higher bandwidth, more densification and cloudification for a hyper connected world by joining billions of devices and nodes with global reach for terrestrial, ocean and space, automated security utilizing the concepts of security function softwarization and virtualization, and machine learning will be inevitable. There are two facets: on the one hand, security algorithms can use machine learning to orchestrate attacks and respond to them in an optimal way. On the other hand, also the attacking algorithms can learn better how the network operates and create better attacks. Continuous deep learning is needed on a packet/byte level and applying machine learning to enforce policies, detect, contain, mitigate and prevent threats or active attacks.

**Research challenge 5: Physical layer security in 6G.** Physical layer security techniques can represent efficient solutions for securing the most critical and less investigated network segments which are the ones between the body sensors and a sink or a hub node. Research questions include which are the most suitable physical layer features to be exploited for the definition of security algorithms in 6G challenging environment characterized by high network scalability, heterogeneous devices and different forms of malicious attacks, and should PhySec be a stand-alone security design or interactions with upper layers are mandatory in 6G networks.

**Research challenge 6: Privacy as exploited resource in 6G.** The relevance specifically for 6G is that, 5G is still largely device / network specific, 6G envisages far more immersive engagement with the network. It is now the subject of ongoing discussion in the standards world. There is currently no way to unambiguously determine when linked, deidentified datasets cross the threshold to become personally identifiable. This is a major, unaddressed problem for many digital technologies in different sectors, such as in Smart Healthcare, Industrial Automation, and Smart Transportation. Courts in different parts of the world are making decisions about whether privacy is being infringed without formal measures of the level of personal information, while companies are seeking new ways to exploit private data to create new business revenues. As solution alternatives, we may consider blockchain, distributed ledger technologies and differential privacy approaches.

## Acknowledgement

This draft white paper has been written by an international expert group, led by the Finnish 6G Flagship program (http://6gflagship.com) at the University of Oulu, within a series of twelve 6G white papers to be published in their final format in June 2020.

## List of contributors

***Editor in chief***
Mika Ylianttila, mika.ylianttila@oulu.fi, Centre for Wireless Communications, University of Oulu, Finland.

***Section editors***
Raimo Kantola, raimo.kantola@aalto.fi, Department of Communications and Networking, Aalto University, Finland
Andrei Gurtov, gurtov@acm.org, Department of Computer and Information Sciences, Linköping University, Sweden
Lozenzo Mucchi, lorenzo.mucchi@unifi.it, Department of Information Engineering, University of Florence, Italy
Ian Oppermann, ianopper@outlook.com, NSW Government Australia, University of Technology Sydney, Australia

***Section contributors***
Zheng Yan, zheng.yan@aalto.fi, Department of Communications and Networking, Aalto University, Finland
Tri Hong Nguyen, tri.nguyen@oulu.fi, Centre for Ubiquitous Computing, University of Oulu, Finland
Fei Liu, liufei19@huawei.com, Singapore Research Center, Huawei International
Tharaka Hewa, tharaka.hewa@oulu.fi, Centre for Wireless Communications, University of Oulu, Finland
Madhusanka Liyanage, madhusanka@ucd.ie, University College Dublin, Ireland
Ahmad Ijaz, ijaz.ahmad@vtt.fi, VTT Technical Research Centre of Finland Ltd, Finland
Juha Partala, juha.partala@oulu.fi, Center for Machine Vision and Signal Analysis, University of Oulu, Finland
Robert Abbas, robert.abbas@mq.edu.au, Macquarie university, Australia
Artur Hecker, artur.hecker@huawei.com, Huawei Technologies Munich Research Center, Germany
Sara Jayousi, sara.jayousi@unifi.it, Department of Information Engineering, University of Florence, Italy
Alessio Martinelli, alessio.martinelli@unifi.it, Department of Information Engineering, University of Florence, Italy
Stefano Caputo, stefano.caputo@unifi.it, Department of Information Engineering, University of Florence, Italy
Jonathan Bechtold, j.bechtold@wiosense.de, WIOsense GmbH & Co. KG, Bremen, Germany
Iván Morales, i.morales@wiosense.de, WIOsense GmbH & Co. KG, Bremen, Germany
Andrei Stoica, r.stoica@wiosense.de, WIOsense GmbH & Co. KG, Bremen, Germany
Giuseppe Abreu, g.abreu@jacobs-university.de, Jacobs University Bremen, Bremen, Germany
Shahriar Shahabuddin, shahriar.shahabuddin@nokia.com, Mobile Networks, Nokia, Oulu, Finland
Erdal Panayirci, eepanay@khas.edu.tr, Kadir Has University, Istanbul, Turkey
Harald Haas, h.haas@ad.ac.uk, University of Edinburgh, UK
Tanesh Kumar, tanesh.kumar@oulu.fi, Centre for Wireless Communications, University of Oulu, Finland
Basak Ozan Ozparlak, basak@ozan.av.tr, Ozyegin University Faculty of Law Istanbul, Turkey
Juha Röning, juha.roning@oulu.fi, Biomimetics and Intelligent Systems Group, University of Oulu, Finland

## Citation information

M. Ylianttila, R. Kantola, A. Gurtov, L. Mucchi, I. Oppermann (eds), "6G White paper: Research challenges for Trust, Security and Privacy". 6G Flagship, University of Oulu, arXiv preprint, April 2020, arXiv:2004.11665, https://arxiv.org/abs/2004.11665

M. Ylianttila, R. Kantola, A. Gurtov, L. Mucchi, I. Oppermann (eds), "6G White paper: Research challenges for Trust, Security and Privacy". 6G Flagship, University of Oulu, June 2020.

# Abstract

The roles of trust, security and privacy are somewhat interconnected, but different facets of next generation networks. The challenges in creating a trustworthy 6G are multidisciplinary spanning technology, regulation, techno-economics, politics and ethics. This white paper addresses their fundamental research challenges in three key areas:

*Trust:* Under the current "open internet" regulation, the telco cloud can be used for trust services only equally for all users. 6G network must support embedded trust for increased level of information security in 6G. Trust modeling, trust policies and trust mechanisms need to be defined. 6G interlinks physical and digital worlds making safety dependent on information security. Therefore, we need trustworthy 6G.

*Security:* In 6G era, the dependence of the economy and societies on IT and the networks will deepen. The role of IT and the networks in national security keeps rising – a continuation of what we see in 5G. The development towards cloud and edge native infrastructures is expected to continue in 6G networks, and we need holistic 6G network security architecture planning. Security automation opens new questions: machine learning can be used to make safer systems, but also more dangerous attacks. Physical layer security techniques can also represent efficient solutions for securing less investigated network segments as first line of defense.

*Privacy:* There is currently no way to unambiguously determine when linked, deidentified datasets cross the threshold to become personally identifiable. This is a major, unaddressed problem for many digital technologies in different sectors. Courts in different parts of the world are making decisions about whether privacy is being infringed without formal measures of the level of personal information, while companies are seeking new ways to exploit private data to create new business revenues. As solution alternatives, we may consider blockchain, distributed ledger technologies and differential privacy approaches.

# 1. Trust networking for 6G

Section editor: Raimo Kantola

Section contributors: Zheng Yan, Tri Hong  Nguyen, Fei Liu, Tharaka Hewa, Madhusanka Liyanage

This chapter discusses the motivations, use cases, solutions, relevant technical mechanisms, visions and research challenges in trust management in beyond 5G and 6G systems.

## 1.1. What is trust networking?

Trust in a network context is about expected outcomes of communicating with a remote party in a session, when clicking on a link or believing in what an email says. The possible outcomes are either the positive value of the communication or being hacked or cheated in some way. Trust spans all protocol layers from the IP layer to applications and content.
A system of trust in a network communication, i.e. trust networking should help in addressing questions like: can this host communicate with a remote party without being attacked or hacked in the process? Can this interaction lead to loss of data? Should flows from a remote network be served or not under heavy load? Or would it be best to drop this flow and devote resources to other flows? Is it possible that a source address in a packet is spoofed? When communicating with a remote party, how does a host minimize its exposure to possible future attacks or long-term loss of privacy? How do the parties protect their communicated data from leakage or being accessed by any unauthorized parties?  In state of the art, a number of frameworks for trust networking can be found: ITU-T Y.3052 [1], Y3053 [2], virtual network and SDN trust framework and Customer Edge Switching (CES) [3]. The latter is advanced since it supports personalized security policies for all devices, at extreme, turning a network into a firewall such that the network transmits only expected traffic [4]. It also supports interworking with legacy IP networks.

**Why is trust networking needed in 6G?**
6G will be used to build a wide digital/physical world boundary for sensing the world, understanding it and programming it. As a result, in addition to loss of information, loss of control over your device or host or loss of money, breach of information security can endanger physical safety of people and cause loss of property. At worst, during international conflicts foreign cyberwar troops could cause havoc in a country on a level that using traditional warfare will not be needed to pressure the victim to accept the terms and conditions issued by the attacker. Like we are seeing in the 5G era, national security concerns are playing an increasing role in mobile technology. This trend will be even more prominent in 6G. To address these concerns 6G network must support embedded trust such that the resulting level of information security in 6G and the packet data networks where 6G provides connectivity to is significantly better than in state-of-the art networks commonly used today.

**Use cases for trust networking in 6G**
Trust networking can be applied in *specialized 6G networks* for local or nation-wide use, it can be applied to packet data networks (PDN) that provide remote access to such specialized networks or any other critical infrastructure. It could be offered to consumers as a PDN service. Finally, trust networking may be applied to all mobile use. The use cases range from simple, single administration handling the whole trusted network with all its devices to cases where devices are owned by independent parties and different segments, layers or subsystems in the network are owned and managed by multiple different stakeholders with possible conflicts of interest. Therefore, trust networking should be able to handle multi-admin relations in flexible ways. Trusted networking shall be provided over multiple trust domains where sharing of trust related information is set up to take place within a trust domain. Some level of trust related information can also be shared across several trust domains. *Trust domain* is a mechanism to scale trust networking into a large number of hosts, network segments, administrations, applications etc.

**Constraints of trust networking**
The introduction of trust networking in an IP network or a mobile network requires changes in the network. It must be possible to deploy those changes one network at a time. If changes are needed both in devices and the network, they are realistically limited to a single administration. An important case of single admin solution is when a server is able to attest the client software and vice versa. If a multi-stakeholder trust networking is desired, no compulsory changes are allowed in the hosts for ease of adoption. We assume that the device market is separate from the network market and that the MNOs cannot dictate security or trust properties of the devices that can be connected beyond what they are able to do today. The current EU "open internet" regulation [8, 9] does not allow personalized filtering of traffic for Internet Access Services (IAS) for consumers by the network. If filtering is applied by the ISP or Mobile Operator (MNO), it must be applied to all users in the same way. While this regulation is in force, trust networking can be applied to *specialized networks* that are out of scope for net neutrality or "open internet". Provided the telco cloud is deregulated, the MNO or ISP could sell trust networking e.g. as "Software as a Service" (SaaS) to its subscribers. The users would then define all aspects of using the cloud software for trust networking and would only use operator defined trust services with a well understood API. This would give the widest impact from trust networking without sacrificing the ideals of "open internet" in any way.

## 1.2. Principles of trust networking

Our earlier White Paper of 2019 [5], ITU-T trusted networking [1,2] and CES [3] all agree on the first principle of applying *ID/Locator split* into the communication such that devices have a stable ID against which all trust and reputation related information can be collected, claims on bad behavior can be made. The edge nodes that assign these IDs will also translate them to IP and other addresses on request. The devices may use private addresses while the edge node will translate between the private addresses and the globally unique routing locators. The ID/Locator split is illustrated in Figure 1. *Policy management* shall be used in trust networking to tailor the generic trust engine to a use case or to end user and admin needs. *Policies* are used to describe the expectations of each entity and many aspects of the behavior of the nodes. Each entity (admin, subscriber, user) shall have its own policy. Policies may scale to the level of individual applications. All aspects of trust negotiation may be controlled by policy.

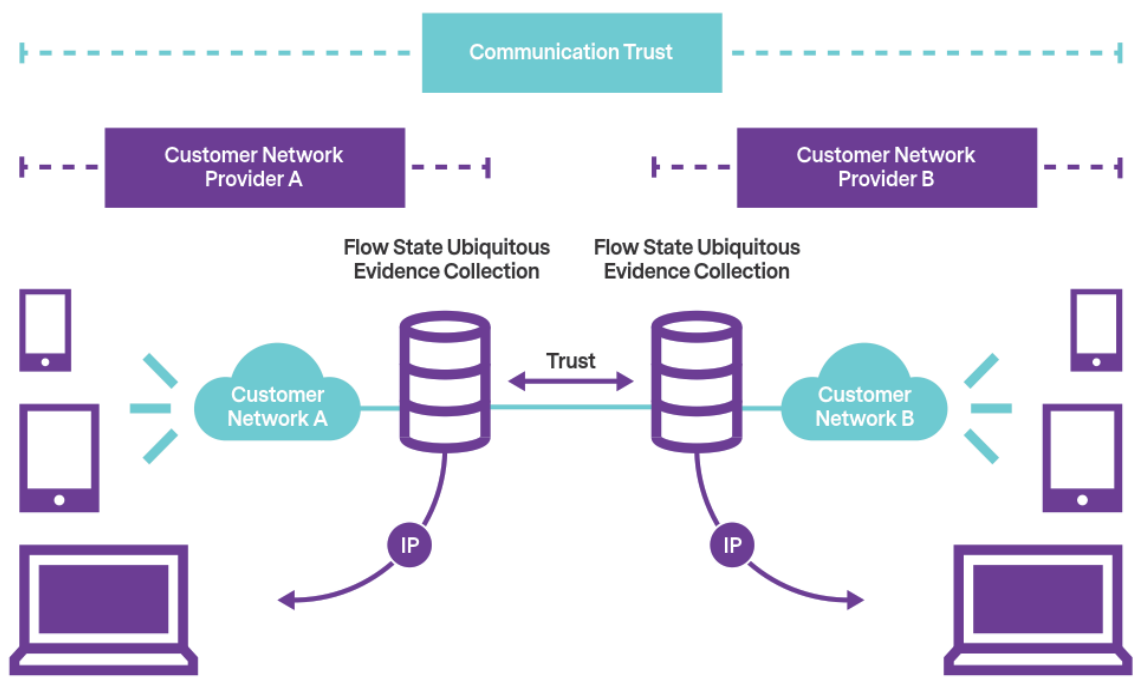

**Figure 1: Conceptual model and ID/locator split for trust networking [5]**

Nodes supporting trust networking shall *ubiquitously collect evidence of behavior* of all seen remote network entities. The collected *evidence may be shared* between nodes in a domain. The evidence is used to produce a *reputation* for the remote entities. The trust networking nodes will have an *embedded reputation system*. Reputation is used to make trust decisions such as asking for more information, admitting a flow, refusing to communicate, allocating resources for an incoming flow depending on the load situation etc. In wide area networking, trust and reputation management may be organized in different ways. The first aspect is to control *trust related information sharing*. Another aspect is how to manage the *trust claims* between the entities and domains such as: "your host Z is too aggressive, restrain it using X", or the explicit acknowledgement like "I have now restored Z into a normal behavior". Alternatively, instead of such an elaborate claims system, reputation may be restored to an indifferent state by *aging*. Wide area trust management might be implemented with a hierarchical or centralized system [6] or by a distributed system or a combination of the two. In order to encourage evidence sharing across administrative boundaries, the evidence could be encrypted and processed in encrypted form by using e.g. homomorphic encryption [7]. In addition, incentives should be provided in order to encourage trust-related data sharing for reputation generation, as depicted in Figure 2.

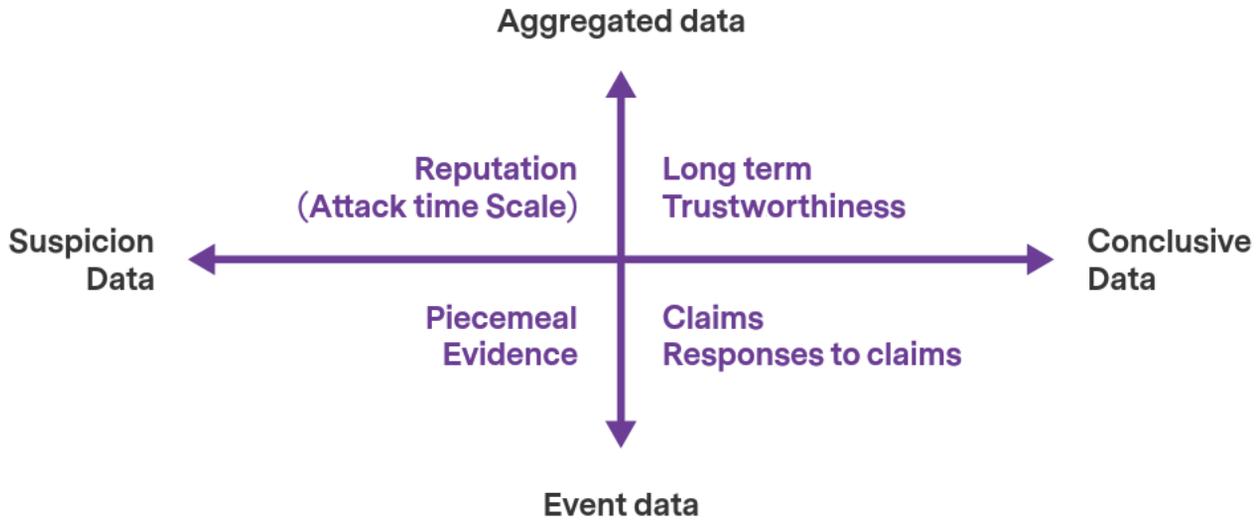

Figure 2: Trust-related data sharing for reputation generation

**Trust modeling**
A trust model describes what evidence is used, how this data is collected, processed, stored and distributed among the stakeholders and how trust decisions are made. The model describes the claims and actions the stakeholders can and will take to restore trust. The model also describes the lifecycle of the trust status of the elements. The myriad of different end user scenarios including vertical networks, advanced vehicular networks, user-tenancy etc., demand that the trust model(s) encompass all elements and entities participating in communications and provide flexibility such that the trust engine is as generic as possible and that it can be tuned to different scenarios and use cases with little effort.

The trust model must have resistance to system attacks such as ballot stuffing or bad-mouthing. To incentivize evidence sharing, data may be shared and processed in an encrypted form. The system should be able to process non-conclusive evidence but all available techniques for producing conclusive evidence should be applied. The trust systems need to be measured and evaluated against the requirements of multiple operators across the globe and the requirements of the regulatory standards ratified across different regions, such as GDPR. In the state of the art, the 5G trust model is focused on the subscriber to MNO relation and on the relations of the different Virtualized Network Functions (VNFs). In Beyond 5G and 6G, an extended trust model for distributed processing including device to device relations should be added in order to support trustworthy use of the mobile system. Trust modeling includes also application trust. This can be supported by trust intelligence, malware detection, software attestation etc. There is a need that the 6G trust model and industry specific trust models should interwork, in particular when 6G is used on the physical/digital world boundary and the operation is safety critical.

**Constraints on trust models:** Agreement on unified terms of quantifying trust among various telecom institutions as well as governments will impose a new challenge. For example, we may need to define liabilities in case when someone has been careless or malicious and his/her resources are used in attacks against other entities. On the same lines the agreement of security transparency or anonymity/obscurity among these parties will need clear practices and rules.

**Distributed Ledgers and 6G**
On the open Internet, decentralized systems where the participants freely leave and join to be a part of a distributed system has been developing without any controllers. This leaves the issue of trust to the end systems. To verify the trust on the services or participants, a traditional solution is based on a third party who proves and confirms for the correctness of the services. In many cases, the parties cannot agree on a single 3rd trusted party, and in fact many 3rd parties are used in parallel weakening security overall. Now, an alternative is emerging with the use of distributed ledgers (DL) with many alternative consensus mechanisms which can support consensus on trust among all parties in a system. The information stored in DL can be plain text or encrypted. The DL is at best when conclusive facts are stored immutably. In addition, smart contracts can be applied to offer autonomic trust networking by automatically triggering trust networking functions based on evaluated trust relationships. In 6G, when conclusive data related to dynamic trust relations is generated, it could be stored in a DL and the possible conclusive responses of the related parties could be suitably processed. As a result, a long-term history reflecting long term quality or trustworthiness of the different parties could be produced. This could be utilized to drive longer term improvement in the operations that use 6G.

**Trust and routing**

The current Internet wide area routing based on the Border Gateway Protocol (BGP) to which mobile systems are exposed in the connected Packet Data Networks (PDN) has many weaknesses, such as the routing prefixes are sometimes hijacked either due to configuration errors or by malicious intent, convergence can take a long time leading to limited availability of the carried services, Quality of Services over the wide area remains an unsolved problem etc. In order to fully leverage the trust and security solutions developed for 6G itself, it is worth to point out that a more trustworthy routing solution is needed for the Internet and the PDNs to which 6G connect.

## 1.3. Research challenges in trust networking

Many research challenges in trust networking remain.
- Open Internet regulation that allows the widest use of trust networking to the maximum benefit for the end users.
- Verification of trust networking in multiple use cases with various needs.
- Scalability of the policy management to all kinds of devices and application use patterns.
- Scaling trust networking to wide area with multiple stakeholders; Scaling trust networking to high network speeds, large number of flow setups, to large number of remote entities, etc.
- Trust management for the wide area across multiple trust domains.
- Extending trust networking to span the whole end to end session, device to network to device.
- How to improve the privacy of DL in 6G era, but still keep its characteristics?
- A specific DL for 6G mobile network
- A study on particular consensus algorithms for trust networking
- What are the key requirements for meeting the trust system of data transparency, AI anonymization, and privacy protection?
- What does the trust model/set of trust models look like, based on the comprehensive consideration of different phases of the trust lifecycle, different service scenarios, and different roles of the ecosystem?
- What are the main factors related to the initial establishment of trust, trust measurement, and trust decision-making within the network and among the different roles of telecom?

# 2. Network security architecture and cryptographic technologies reaching for post-quantum era

Section editor: Andrei Gurtov

Section contributors: Ahmad Ijaz, Juha Partala, Robert Abbas, Artur Hecker

This chapter discusses about the challenges, solutions and visions of network security architecture and of cryptographic technologies in beyond 5G systems from several aspects. Future of SIM cards and use of asymmetric cryptography are pondered upon. We discuss also future convergence of telecom networks with the Internet infrastructure. Other topics include security of software defined networking with AI capabilities and trustworthy cloud computing with remote attestation when applied to 6G.

## 2.1. Network Security Architecture in 6G

Since inception of digital mobile communication in 2G, mobile networks are reliant on a physical storage of symmetric keys in a Subscriber Identity Module known also as SIM card. Encryption algorithms migrated from customary to international standards, and additional cryptographic mechanisms were added for mutual authentication. However, fundamentally the security model in 5G is still reliant on SIM cards [16]. While SIM cards became smaller (now at "nano" size), those still need to be plugged to devices, which limits applicability e.g. to IoT. Introduction of eSIMs partly addresses this challenge, though leaves issues with physical size. iSIM under development could be a part of System-on-Chip in future devices although it faces opposition from operators due to possible loss of control.

**The need for 6G network security architecture model**
Traditional SIM cards rely on proven symmetric key encryption, which scaled well up to billions of users. But it has drawback for example with IoT, privacy, network authentication and false base stations. Is it going to be a fundamental shift from symmetric crypto to asymmetric public/private keys? This was never deployed before at such scale. In addition to SIM, 5G plans to support authentication through a public-key infrastructure (PKI). The core of 5G is implemented as a set of microservices communicating over HTTPS. The authentication, confidentiality and integrity for such communication is provided by Transport Layer Security (TLS) using elliptic curve cryptography (ECC). However, this has not yet been deployed yet and can be left for 6G.

This chapter tries to answer key questions about 6G security model. Is it still going to be physical SIM cards in the devices? Or most IoT devices will have clones of software SIMs or Trusted Platform Modules? A certificate system for WWW works although with eventual certificate revocation and Certificate Authority (CA) break-ins. DNSSEC is an example of gradually deploying asymmetric key system. Host Identity Protocol implements this concept at the network layer [25]. Preventing Man-in-the-Middle attacks is a critical requirement for asymmetric encryption. Network slicing in 5G, as it has been defined by the 3GPP in Rel15, has hardly any security impact, as it does not permit to separate traffic of different services. Instead, deploying IPsec/VPN or HIP services in 6G would help to isolate user traffic.

## 2.2. Post-Quantum Crypto-Security in the 6G Architecture

The quantum computing paradigm is fundamentally different from classical computing. There are computational problems we do not know how to solve efficiently on a contemporary computer, but there are algorithms that solve those problems efficiently on a quantum one. One of these problems is the discrete logarithm problem which is the basis of modern asymmetric cryptography. If large-scale quantum computing becomes a reality, these cryptographic primitives need to be replaced for quantum-secure ones. According to a recent survey [17], quantum computing may be commercially available in a few years. While large-scale quantum computing can be expected to take longer, it is time to prepare for the shift to cryptography that is secure in the post-quantum world. According to current knowledge, contemporary symmetric cryptography remains secure for the most part even after the advent of quantum computing. In general, it suffices to double the size of the symmetric keys due to Grover's algorithm. The problems lie in asymmetric primitives based on integer factorization and the discrete logarithm problem that are solvable in polynomial time on a quantum computer using Shor's algorithm.

**Why is consideration of quantum computing important for 6G?**

The current 5G standard does not address the issue of quantum computing, but relies on traditional cryptography such as ECC. However, the elliptic curve discrete logarithm problem (ECDLP) can be solved in polynomial time on a quantum computer. The development towards cloud and edge native infrastructures is expected to continue in 6G networks. Compared to earlier generations, the security architecture of 6G will be more complex, dominated by current transport layer security standards and be increasingly dependent on the PKI. This development will make the core network completely reliant on the functionality and security of the underlying PKI. However, currently there are no post-quantum secure primitives, for example, in TLS.

There are public-key primitives considered to be quantum-safe [19]. These include for example, code-based encryption schemes such as McEliece [22] and lattice-based NTRU [21]. Many of these suggestions have survived decades of attacks and can thus be considered secure both in the classical and the quantum setting. However, their efficiency is poor and key sizes big compared to, for example, ECDLP-based schemes. Replacement of contemporary asymmetric cryptography with post-quantum secure schemes will incur costs both in the communication and operational efficiency of the network. Research is needed to identify the correct application of post-quantum secure cryptography in order to satisfy the envisioned performance and functionality of the 6G architecture. In addition, research into new, more efficient post-quantum secure asymmetric schemes is needed in order to reach this goal [18-22]

Standardization efforts for post-quantum cryptography are ongoing. In the United States, National Institute of Standards and Technology (NIST) is currently hosting a selection process NIST PQC for post-quantum cryptography standardization. These new primitives are expected to provide post-quantum secure key exchange, as well as to augment the Digital Signature Standard (DSS) FIPS 186-4. We propose that the 6G standardization community pays close attention to these efforts.

## 2.3. Software and AI defined security in beyond 5G and 6G

As 6G moves toward THz spectrum with much higher bandwidth, more densification and cloudification for a hyper connected world by joining billions of devices and nodes with global reach for terrestrial, ocean and space, automated security utilizing the concepts of security function softwarization and virtualization, and machine learning will be inevitable. To eliminate constraints in existing and evolving 5G networks security, security systems using the existing concepts of SDN and NFV must be further improved with embedding intelligence for dynamicity to match the needs of 6G security. In this vain, intelligent security functions in containerised VNF box will monitor traffic in 6G residing in gateways to scan the traffic using continuous deep learning on a packet/byte level and applying machine learning to enforce policies, detect, contain, mitigate and prevent threats or active attacks. Security functions using container technology offers better utilization rates, less storage requirements, enhanced security and faster reboot time. Containers will be grouped into Pods, each Pod consisting of multiple containers on a single machine, with security service functions and providing availability through scaling up or down. The advances in cloud computing such as edge and fog computing will be used to maintain and deploy security functions (security VNFs) in different network perimeters as the use arises through proactive decision-making using machine learning. Building on the concepts of SDN, global resource visibility and event monitoring, with synchronized network security policies among different stakeholders, and programmable APIs, network abstractions will be used to ensure end-to-end network security.

6G networks will harmonize the concepts of SDN, NFV, and AI in an integrated environment not only to provide the necessary service [23], but also to ensure end-to-end network security, as shown in Figure 3. Programmable interfaces on programmable forwarding plane will enable deploying softwarized security functions much like VNFs in any network perimeter or instance in a virtual environment using AI not only proactively discover threats, but also to initiate security function transfer from point to point throughout the network.

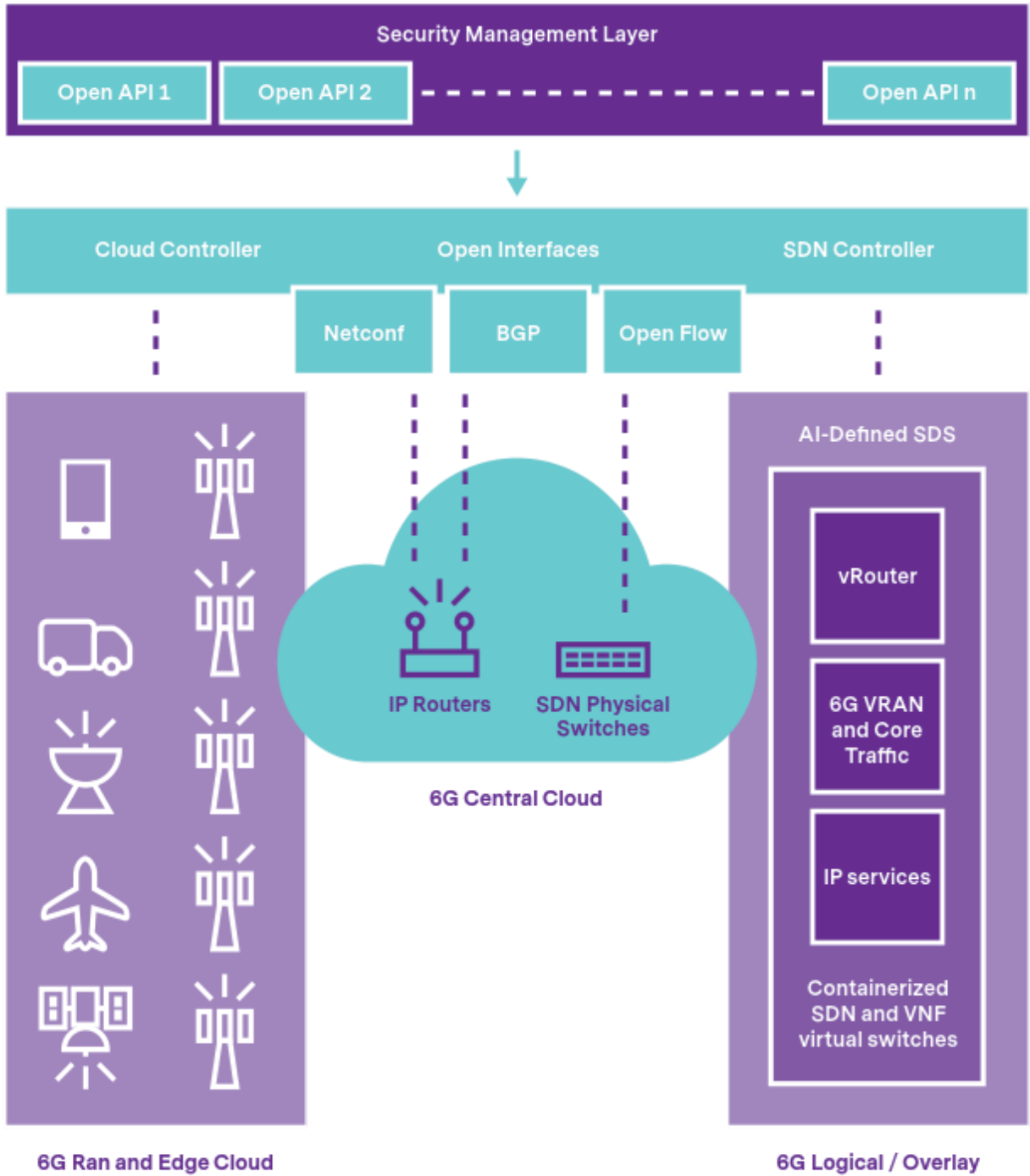

**Figure 3: A holistic software defined security platform leveraging AI**

6G total automated, zero touch and zero trust security where zero trust for all North-South (N-S outbound/Inbound traffic from/to data center) and East-West (E-W form Container to another) cloud traffic must be checked with AI based ML for threat detection [24], prevention and containment where the network can be treated as a giant firewall that integrates the flowing security functions:

- Access security (inbound/ outbound gateway, end points security)
- Cloud security (edge cloud and central cloud)
- Orchestration platform security

## 2.4. Securing the convergence of telco cloud

With the advent of virtualization and softwarization, the Cloud Computing has become the driving paradigm for service provisioning in the smartphone era. With 5G, these principles have found their way both into the recent releases of the 3GPP standards (e.g. Rel15 SBA) and in the underlying support systems (ETSI NFV, ONF SDN, ETSI MEC, orchestration, etc). Furthermore, similar virtualization, sandboxing and Cloud-provided service concepts are widespread in the application areas on the modern powerful terminals: for instance, it is quite common to seamlessly integrate remote (e.g. Cloud-based) service end-points in the smartphone apps; specific computation-intensive or dataset-based functions like optimizations, search, pattern recognition in multimedia, etc., are often seamlessly performed on a remote Cloud.

With the inclusion and the expected further fusion of IoT and mobile telecommunications domains, the number of connected devices, their density and their service needs will increase by several magnitudes, in turn requiring a considerable increase in available network access points, network capacity and service capabilities. Consequently, and given the performance constraints of the typical IoT devices, such offloading of compute, storage and networking capabilities to other nodes can be expected to increase with the success and during the course of 5G and to generalize and finally become common place functionality in 6G. Indeed, given its presumed better reliability and higher throughputs with lower latencies, it is only natural to expect diverse computations in 6G to be offloaded both from terminals to the network (e.g. IoT) and from the network to particular terminals or groups of the latter (Customer-Premises Data Centers in the factories, smart cities and smart hospitals).

**Opportunities related to Execution Offloading in 6G**

If the questions above can be addressed, the generic offloading of computation paves path to also improving the security posture of the system:
- Privacy improvement: users could avoid giving out raw data by only allowing critical computations within their own virtual machines running on their or on remote equipment.
- Green computing: through the support of load-dependent offloading, green computations e.g. on nodes with available eco-produced current become possible. At the system level, it allows to remove nodes such as Data Centers, and therefore to flatten the regional peaks in the power consumption.

**Challenges related to Execution Offloading in 6G**
Executing various functions on different remote ends is an outstanding feature, which allows to dynamically adjust the application to the available resources and the resources to the application requirements and has the potential to introduce new appealing features into 6G. However, it also poses several well-known security threats:
- *Data confidentiality:* if data must be sent to remote endpoints, how can a tenant be sure that this data is only used for the intended computation?
- *Data integrity:* how can we ensure that the data is not lost, modified in the process? This, in particular, applies to the obtained results. How do we know that the communicated results are indeed the expected outcome?
- *Platform integrity:* as such computational offloading can be seen as a generalization of the Cloud-computing, the platform of any service environment therefore will spread dynamically over all resources currently involved into the computation. Therefore, in addition to the questions above, the question of the integrity of that virtual, possibly dynamically changing, service computation domain should be addressed.
- *User privacy:* in addition to the raw data, which might or might not be private data of the user, the statistics and meta-data of the requests should be protected against abuse.

**Existing Approaches and Research Challenges**
There are several approaches, how the security challenges above can be overcome or mitigated:
- *Authentication of remote endpoints:* authentication of endpoints allows to verify the claimed identity of the endpoint. However, it is not capable of making statements with regard to the integrity of that endpoint. Therefore, it is not alone sufficient to protect against such threats as errors in the internal implementation, failures during the runtime, malicious modifications of the remote endpoint (e.g. after intrusion).
- *Certification of the platforms:* certification of the platform verifies particular features and claims with regard to that platform. Security verification methods such as Common Criteria are well-understood and could be applied to verify the soundness of the implementation of a set of security mechanisms (e.g. as per protection profile). However, such measures are also costly and cannot per se guarantee the security posture of a node at a particular point of time, as intrusions cannot be prevented.

- *Support of remote attestation:* some security frameworks, such as the so-called Trusted Computing (TC), can provide verifiable remote attestation services in some scenarios. By tying key platform operations to tamper-resistant on-board hardware modules (so-called TPM) and introducing a cryptographic framework on top of the latter, the integrity of a remote platform including its operating system, virtual machines and services can be addressed. However, TC also requires a sound key management and standardized interfaces. Generally speaking, TC raises many questions as to the feasibility (e.g., scalability) and economic viability (e.g. OPEX overhead) of such integration in the operational highly multi-tenant environment of 6G (with many operators, verticals and end-users).
- *Support of secure properties in spite insecure execution platforms:* in the recent years, quite some research has been devoted to the novel mechanisms, which do not require a trusted platform for (particular) secure computations. Among such approaches, we can broadly mention Encrypted Search, Private Information Retrieval and Search, ORAM, Fully Homomorphic Encryption libraries, etc. Since these methods do not require platform certification, they have the potential for better inclusion (also on the fly) and for cost reduction. However, not all of these methods are suitable for general offloading, and some are known to be extremely computation-intensive. Therefore, it remains to be seen, if, where and how such and similar methods can be integrated in the 6G.

The list above is not mutually exclusive and several such mechanisms can coexist. The impacts of such co-existence should be carefully addressed in the operational reality of mobile networks.

## 2.5. Research challenges for 6G security architectures

Among important general challenges is whether future 6G networks will reply on web CAs or DNSSEC instead of SIM cards to admit devices to the network. Another fundamental issue is legal interception requirement for current telecom operators providing mobile services. It is going to be still present in 6G with all-IP voice, chat, and data communication? Are authorities required e.g. to read all sensor, remote vehicle or tactile Internet data under a court order? This affects fundamentally the security architecture, as end-to-end encryption would not be allowed in place of proxy-time communication with key escrow.

**Research challenges/questions:**

- How to replace SIM cards for IoT devices?
- How to apply 6G security automation vision based on full visibility?
- How to utilize AI to provide real time full protection End to End against known and unknown threat?
- 6G Security platform and framework.
- How to ensure post-quantum security of 6G and how will post-quantum secure cryptography affect the performance of 6G networks?
- What post-quantum cryptographic primitives will be the optimal choices for 6G?
- Secure remote computations and secure offloading

# 3. Physical Layer Security solutions and technologies

Section editor: Lorenzo Mucchi

Section contributors: Sara Jayousi, Alessio Martinelli, Stefano Caputo, Jonathan Bechtold, Iván Morales, Andrei Stoica, Giuseppe Abreu, Shahriar Shahabuddin, Erdal Panayirci, Harald Haas

6G is envisioned to be a full connectivity fabric, whose nodes can span from satellite to inside the human body. This ultra-dense network of heterogeneous nodes provides tons of information, often extremely sensitive. In this context, full security is a mandatory feature to let people trust the services. Physical-layer security is the first line of defense, and it can provide security even to low complex nodes in different scenarios. This chapter discusses about the challenges, solutions and visions of physical-layer security in beyond-5G systems from several aspects. Section 3.1 discusses the vision of PhySec over 6G and provides, as an example of application, the human body communications. Section 3.2 discusses the use of PhySec for key distribution, attack detection and protocols for low-complex D2D communications. Sections 3.3 and 3.4 provide effective implementations of PhySec, from massive MIMO and intelligent reflecting surfaces to visible-light communications.

## 3.1. Physical-layer security as confidentiality enabler in 6G connectivity

Moving beyond the 5G technology, 6G will enhance the key performance indicators of 5G, enabling the definition of more demanding applications, ranging from augmented reality and holographic projection to ultra-sensitive applications. In this context, a holistic approach of security is required to cope with the plethora of different systems and platforms. The large amount of the world data collected by networks of sensors (environmental, human-body, etc.) and the mobility features of most scenarios ask for advanced security techniques that take into account new constraints in terms of device capabilities, network environment and network dynamic topology [26]. Physical layer security, moving the security strategy at physical layer, might be one of the confidentiality enablers in 6G connectivity. Its features, combined with the advances in artificial intelligence algorithms and the trend of distributed computing architectures, can be exploited either to enhance the classical cryptographic techniques or to meet the security requirements when dealing with simple but sensitive devices which are unable to implement cryptographic methods, e.g. devices and nano-devices of the internet of things and bio-nano-things where the human inner bodies become nodes of the future internet [27] (Figure 4).

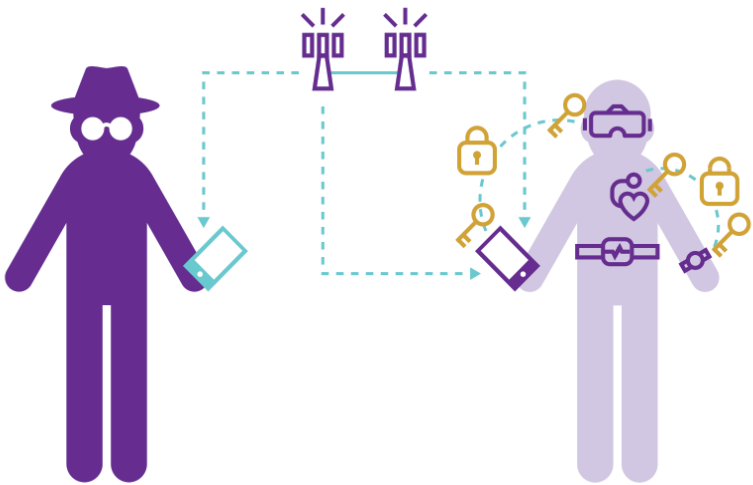

**Figure 4: 6G needs to solve novel physical layer security threats**

Computational and energy resources of a network node can be reduced by adapting the security algorithm to the environmental context where the communication occurs, leading to the definition of a context-aware security approach. The dynamic context in terms of mobility, network nodes density, frequency spectrum utilization and technology heterogeneity which is envisaged in 6G scenarios should be taken into account in the definition of security communication strategies both for the identification of the level of security countermeasure needed in a specific moment and for the exploitation of these environmental characteristics in the security algorithm definition. Environmental and operational intelligent physical layer security also based on the adoption of Artificial Intelligence algorithms may lead to a the definition of new techniques that can early detect the need of enhanced security mechanism to be dynamically activated (e.g. based on the battery level of the involved devices or the degree of trustworthiness of the specific context) and do not considerably affect the transmission spectral efficiency [28]. This approach complies with the main 6G key features that the enabling communication technologies should meet in term of low energy consumption and long battery life, high affordability and full customization and distributed artificial intelligence architectures.

Physical layer security addresses one of the most important application of 6G: the human-centric mobile communications. In this framework, an increasing interest of scientific research has been oriented to wireless body area network and in particular to on-body and in-body nano-devices, including biochemical communications. In the next future, the human body will be part of the network architectures, it will be seen as a node of the network or a set of nodes (wearable devices, implantable sensors, nano-devices, etc.) that collect sensitive information to be exchanged for multiple purposes (e.g. health, statistics, safety, etc.). By coping with the high security and privacy requirements and the energy and miniaturization constraints of the new communication terminals the Physical layer security techniques can represent efficient solutions for securing the most critical and less investigated network segments which are the ones between the body sensors and a sink or a hub node.

Two interesting potential application scenarios for physical layer security in 6G context are Human Bond Communication and Molecular Communication. The former requires a secure transmission of all the five human senses for replicating human biological features, allowing disease diagnosis, emotion detection, biological characteristics gathering and human body remote interaction. While the latter, based on the shifting of the information theory concepts in the biochemical domain (communications among biological cells inside the human body) requires advanced low-complexity and reliable mechanisms for securing intra-body communications and enabling trustworthy sensing and actuation in a challenging environment as the human body is (e.g. secure Internet of bio-nano things) [29]. ETSI SmartBAN group is working on the standardization of security & privacy for the future body area networks, and physical layer security is one candidate to handle the confidentiality of in- and on-body devices with typically low resources available. This is important also when 6G will include in- or on-body nodes as part of the Network (Figure 5).

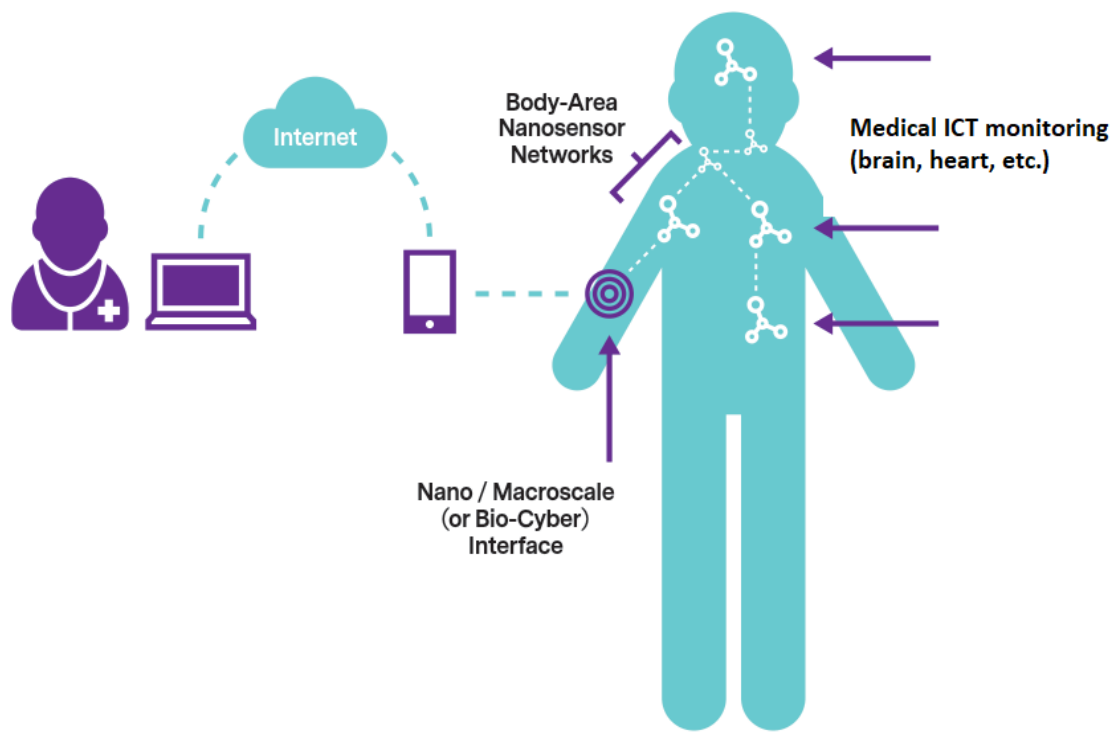

**Figure 5: Human body as part of the global network.**

## 3.2. Distributed and cooperative PHY-security protocols for 6G networks

Besides providing a keyless and innately secure communication channel via maximization of secrecy rate, PLS may also exploit the intrinsic characteristics of the wireless channel to co-generate a cryptographic key for symmetric encryption. For latency-constrained communication scenarios and resource-constrained radio devices the secrecy enhancing techniques detailed above become cumbersome or impractical. This is usually the case for high device densities under opportunistic self-organizing network formation paradigms. State-of-the-art encryption itself is considered unassailable when it comes to data confidentiality and integrity, however, there exists doubts in the traditional authentication and key distribution design for the future.

PHY-based key generation solutions distinguish themselves from traditional key exchange solutions by being completely decentralized and not relying on any fixed parameters designed by a particular entity, but rather on the distributed entropy source that is the wireless channel. So far, in D2D protocols the totality of the raw data needed to synthesize such a key has not been readily available to higher abstraction levels. In future, newly developed or extended communication protocol implementations should see PHY-layer attributes (CSI, RSSI, CFO, etc.) of all PHY-exchanges easily available to higher layers, allowing for a much deeper level of integration, control, and interchangeability of security modules. Such attribute granularity and unprecedented network visibility at the PHY-level will encourage the development of security and authentication solutions which leverage these previously unused characteristics. Thus, providing resilience to the existing vulnerabilities that the state-of-the-art Diffie-Hellman (DH) key exchange algorithm poses [30] as well as provide immunity against the real-time computation of discrete logarithms that will come about by the time 6G is deployed and quantum computing has matured.

Besides current vulnerabilities at a cryptographic level, more issues arise with future network deployments thanks to the introduction of D2D communication in 3GPP Release 12 [31], opening the door to Proximity-based services (ProSe) [32]. If a comprehensive security is not embedded into these services, new attack vectors will arise at the PHY-layer in the form of range extension/reduction attacks that can spoof distances between devices. With a secure communication link, the available PHY-layer attributes can then also be used for PHY-threat detection. These can be implemented using classical signal processing techniques that highlight and discover anomalies in the PHY-attributes of the particular received signal or through the detection of abnormalities in the packet exchanges.

Such lightweight implementations are ideal for networks of resource-constrained devices; however, the true potential for threat detection lies in learning (ML), where massively aggregated attributes can be used to train the ML models for monitoring and classification. The usage of ML methods in networks with high PHY-Attribute visibility will enable real-time PHY-Layer monitoring and knowledge-based detection, making it highly attractive for leading AI companies to develop Security-as-a-Service (SecaaS) applications. The deployment of SecaaS applications will be highly dependent on the topology of the networks to be safeguarded. For instance, in networks with facilitator nodes (and/or gateways) which actively route local packets, such nodes can be used as data aggregators of the necessary PHY-attributes of communicating devices and their respective links for active threat detection. Alternatively, a passive observer with higher processing power can be introduced as part of a SecaaS application, functioning as the aggregator of the exchanged packets to enhance the embedded D2D threat real time detection capabilities of the network.

Eventually a security paradigm will form in which a great number of highly diversified, independently generated threat detection models will be formed at each aggregator/node. This diversity of models can then open up the possibility for the deployment of transfer learning techniques to share learned parameters between adjacent networks in order to detect novel malicious attacks and prepare networks against attacks like a vaccine would.

### 3.3. Security of Cell-free Massive MIMO and Intelligent Reflective Surface

The two physical layer technologies that grabbed most attention from the research community are: (1) Cell-free massive MIMO and (2) Intelligent reflective surface (IRS). They are currently the two strongest candidates for physical layer of 6G communication systems.

Cell Free Massive MIMO. A large number of antennas are typically used to equip a bulky and expensive massive MIMO BS which resides on an elevated location, for example, on top of a building, to increase the size of cell radius. A single massive MIMO BS covers a large number of users from a distance and therefore, large variations of received signal strength exists between different users [33]. Cell-free massive MIMO is a form of network MIMO where the antennas are not centralized but distributed among different locations. A central baseband processing unit, which is connected to all the antenna stripes through cables, is used to perform the necessary baseband signal processing operations [34]. The antenna stripes can be as small as a matchbox in size and can be integrated in an adhesive tape as displayed by Ericsson at MWC 2019.

The biggest security issue of a cell-free massive MIMO system is the exposed location of radio stripes. A local active attacker may get physical access to the BS and interfere with the internal elements. The attacker may exploit internal lines by direct wiretapping to inject malicious software and configuration parameters. Due to their vulnerable location, passive attacks, such as, eavesdropping on authentication keys, user-specific keys and short-term session keys may also become easier. It is rather easy to destroy a miniature stripe from their exposed physical location than a bulky massive MIMO BS to disrupt the overall communication. In addition, the use of complex encryption method, to provide data confidentiality between antenna stripes and central baseband processing unit, is also not possible due to the small size of an antenna stripe.

Intelligent Reflective Surface (IRS) is comprised of an array of IRS units which can be used to change the phase, amplitude, or frequency of incident signals. Typically, signals transmitted from different antennas are sent towards IRS, which reflects a beamformed signal towards the legitimate users (Figure 6). Thus, IRS creates an alternative transmission path when the line-of-sight (LOS) is blocked between the transmitter and receiver. The IRS scheme will be particularly important for high frequency communications where the penetration loss is significant [35]. Massive MIMO uses techniques like beamforming and jamming with artificial noise insertion to secure physical layer communications.

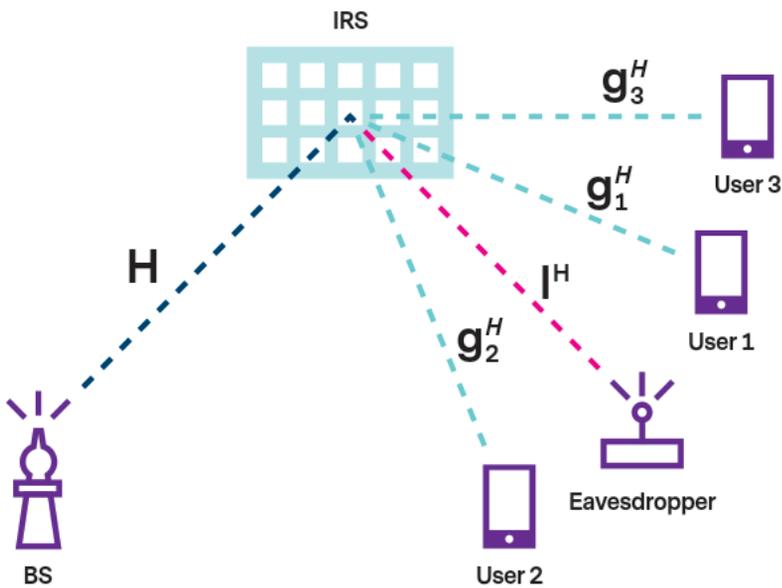

**Figure 6: MIMO and IRS focusing information to legitimate users while noise-only to eavesdroppers.**

However, the achievable secrecy rate is limited even with these techniques when the links of the legitimate user and the eavesdropper are highly correlated. IRS can be used in such a scenario to constructively add the beamformed signal towards the user and destructively add towards the eavesdropper. As the signal travels in a NLOS path, it is difficult for the eavesdropper to detect the incident angle of the signal. However, to make the system secure with IRS, the system needs to detect and locate the eavesdropper which is not a trivial issue. The IRS controller, which controls the phase of IRS, can be compromised by an active attacker to focus the beam towards unintended users. If the location of the IRS is exposed, a passive attacker can also locate itself near the IRS to exploit a correlated channel for eavesdropping.

### 3.4. Physical Layer Security using Visible Light Communications in 6G

The area of physical layer security (PLS) can play a vital role in reducing both the latency as well as the complexity of novel security standards. It is expected that the dramatic increase in high data rate services will continue its trend to meet the demands of 6G networks. Optical Wireless Communications and one of its variants, Visible Light Communications (VLC), offer attractive features such as high capacity, robustness to electromagnetic interference, a high degree of spatial confinement, inherent security (Fig. 7) and unlicensed spectrum. VLC is particularly considered as an emerging technology that has been introduced as a promising solution for 6G. It is a special form of optical wireless communications and uses white-LEDs to encode data in the optical frequencies and some works have suggested that it is a good candidate to meet the data rate requirements of 6G. Depending on the intended application, VLC can serve as a powerful complementary technology to the existing ones. Such as, Wireless Body Area Network (WBAN) and personal area network (PAN), Wireless Local Area Network (WLAN), Vehicular Area Network (VANET) and Underwater hybrid acoustic/VLC underwater sensor network. VLC will also be useful in scenarios in which traditional RF communication is less effective such as in-cabin internet service in airplanes, underwater communication, healthcare zones and etc.

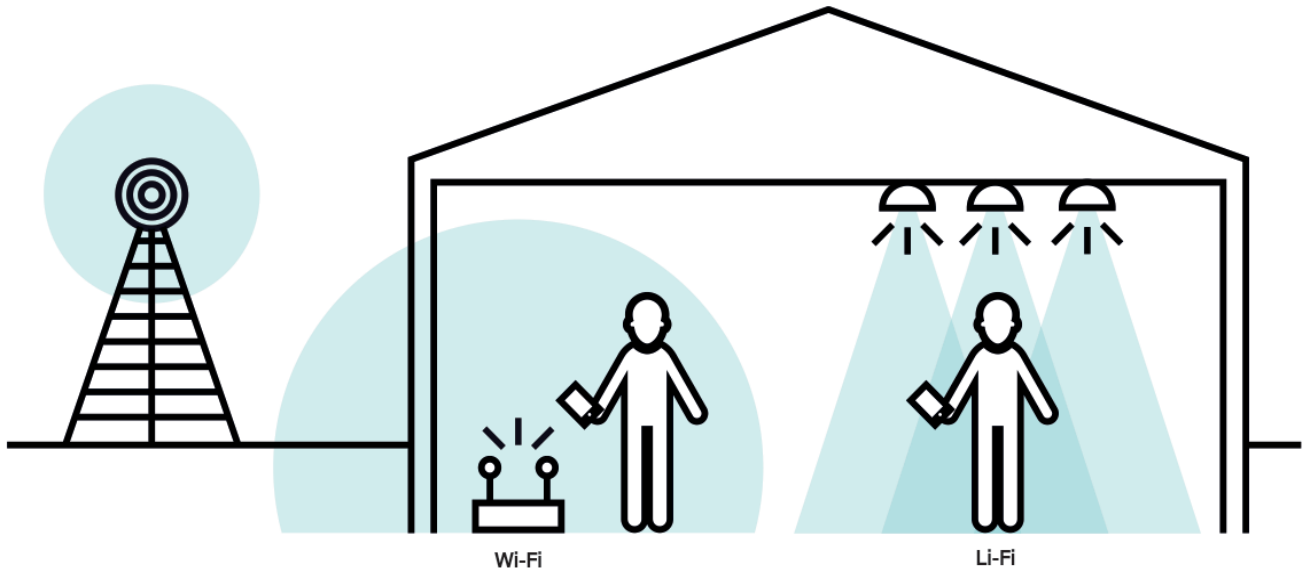

**Figure 7: Visible light communications (Li-Fi) vs radio frequency communications (Wi-Fi).**

During the past few years, PLS in VLC networks has emerged as a promising approach to complement conventional encryption techniques and provide a first line of defense against eavesdropping attacks. The key idea behind it is to utilize the intrinsic properties of the VLC channel to realize enhanced physical layer security. The evolution toward 6G wireless communications poses new and technical challenges which remain unresolved for PLS in VLC research, including physical layer security coding, massive multiple-input multiple-output, non-orthogonal multiple access, full duplex technology and so on. Moreover, it is not possible to employ conventional PLS techniques as in RF communications, in fact the most practical communication scheme for VLC systems is intensity modulation and direct detection. Due to the nature of light, the intensity-modulating data signal must satisfy a positive-valued amplitude constraint. A brief summary of the PLS enhancement methods proposed for VLC is shown in Fig. 8, and the details are given in [36-39].

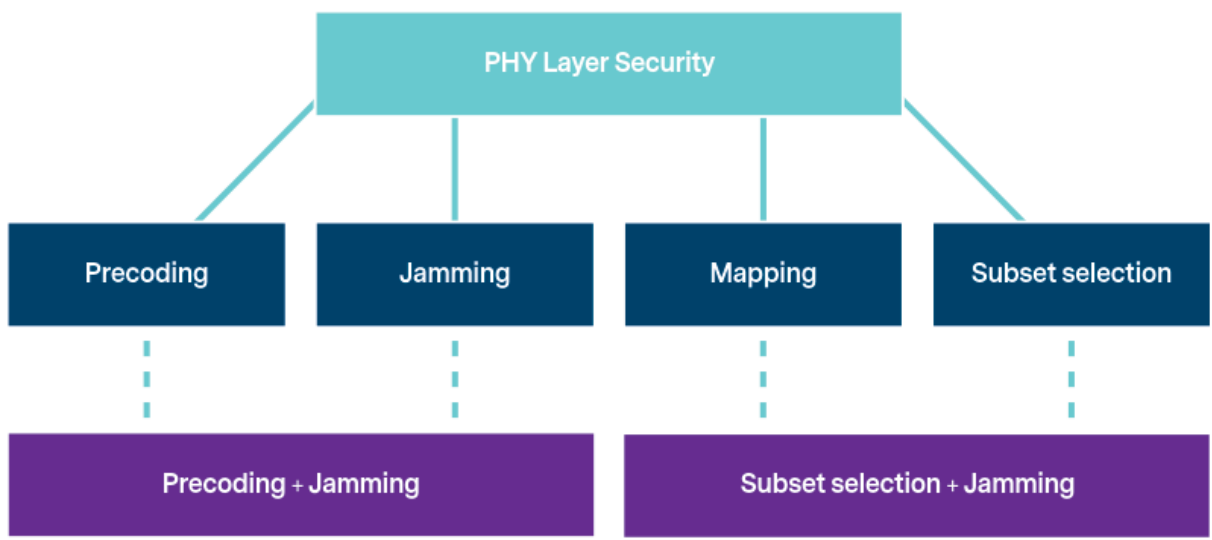

**Figure 8: Taxonomy of the PHY layer security for VLC systems**

## 3.5. Research challenges for physical layer security

- Which are the most suitable physical layer features to be exploited for the definition of security algorithms in 6G challenging heterogeneous environments characterized by high network scalability and different forms of active malicious attacks?
- How artificial intelligence can be exploited to dynamically tune physical layer security algorithms?
- How to best develop lightweight key distribution and authorization techniques that leverage the previously obscured PHY-layer attributes, while maintaining ULL QoS?
- What are the relevant, unique dimension reduction / feature extraction methods to enable transfer learning while maintaining the privacy of aggregator networks over various RF interfaces?
- How to ensure platform security with antennas distributed in different locations?
- How to provide confidentiality between central baseband processing unit and antenna stripes?
- What kind of security mechanisms could be used between the transmitter and the IRS panel and how to ensure security of IRS controller?
- New and novel algorithms for PLS in multiuser and broadband VLC systems, with new modulation schemes such as spatial modulations techniques that are derived from it such as index modulation, space shift keying, OFDM-index modulation techniques (OFDM-IM). Other techniques include optical multiple-input-multiple output with non-orthogonal multiple excess (NOMA) system.
- The algorithms to be designed have high power efficient and must have the capability to work in multi-user scenarios. In particular, the artificial jamming signal generation property of these modulation techniques is the most important advantage in providing PLS compared to the traditional approaches. Moreover, the theoretical methods, to develop the maximum achievable secrecy capacity and secrecy rate of the physical layer security algorithms will be much different than the approaches adopted by the by traditional RF systems because of the different system architectures employed.
- Also, incorporating user mobility and device orientation into the VLC channel models and combining VLC and RF systems pose new challenges in PLS research and development.

# 4. Privacy protection in 6G: principles, technologies and regulation

Section editor: Ian Oppermann

Section contributors: Tanesh Kumar, Basak Ozan Ozparlak, Juha Röning

This chapter discusses about the challenges, solutions and visions of privacy protection in beyond 5G systems from several aspects. Developing a measure of Personal Information (PI) in linked identified data sets used to provide smart services, and a threshold test for Personally Identifiable Information (PII) which considers personal features, spatial and temporal aspects of data, a context within which data is captured and analyzed. Use these measures and frameworks to underpin governance systems which support regulatory requirements for the protection of citizen data in different economies which adopt beyond 5G technology. Moreover, this chapter presents some of the potential privacy-preserving technological candidates that may be vital for future wireless applications. At the end, some insight is given about standardization and regulatory aspects in the context of 6G privacy.

## 4.1. Privacy Requirements in a future hyper-connected mobile world

6G is expected to move us much further towards the ideal of ubiquitous connectivity for a myriad of devices, sensors and autonomous applications. This creates the fundamental enabler for Future Smart Services for homes, factories, cities, and governments, which in turn rely on sharing of large volumes of often personal data between individuals and organizations, or between individuals and governments [40]. A smart light in your home which turns on and off as you move around the house can provide a more efficient use of energy for lighting, but will use de-identified data about when you are home, which rooms you use and when, if there are other people in your home, where in your home you spend your time. Within this deidentified data, there are insights about you, your relationships, habits and preferences. In aggregate form, this data can be used by a smart lighting provider to deliver more efficiency lighting services to a suburb, or by a smart grid to match energy demand to energy supply or by a smart micro energy service provider to make best use of spot energy prices.

The benefit is the ability to create locally optimized or individually personalized services based on personal preference, as well as an understanding of the wider network of users and providers. When we consider just how wide scope "smart" covers in our personal lives (smart TV, smart scales, smart toilet, smart phone, virtual assistant, smart home, smart car), or in the wider community (smart grid, smart materials, smart factory, smart city, even smart government), the benefits in terms of improved efficiency, improved effectiveness and increased personalization to our individual needs can be enormous. If these datasets are linked, a great of personal information may be contained in the joined data, sufficient to reasonably reidentify individuals represented in the data. How this data is used and by whom for what purposes creates risks and concerns. In many economies, it may also force services providers and operators to think about difference governance paradigms to support regulatory requirements such as GDPR.

With the multidimensional flows of rich data, the challenge is quantifying what deidentified data means, to develop measures for the level of personal information in a data set at any point in time, and to develop threshold tests for when an individual is reasonably identifiable (PII) all while considering personal attributes, temporal and spatial aspects of data, and rich contextual environments. Furthermore, the post 5G smart ecosystem will be a shared network infrastructure where multiple stakeholders collaboratively provide diverse set of services to the consumers. Therefore, it also opens up the discussion between the trade-off between managing privacy with building the required trust. As we give more trust to the involved entities/stakeholders, the risk of privacy leakage increases. Thus 6G will require new trust models along with updated privacy protection approaches to provide balance between maintaining consumer's privacy and trust [41, 42].

## 4.2. How Personal is Information?

The terms personal information (PI) and personally identifiable information (PII) are often used interchangeably in legislative frameworks as well as in different technical literature. PI is typically described in a way that covers a very wide field, and this description varies in different parts of the world. For example [43]:

"… personal information means information or an opinion (including information or an opinion forming part of a database and whether or not recorded in a material form) about an individual whose identity is apparent or can reasonably be ascertained from the information or opinion" [2 Section 4].

For example, date of birth is considered PI but not PII as it cannot uniquely identify an individual. Similarly, spatial location or time of service use will not uniquely identify anyone, except in isolated cases. The question becomes, how many features must be linked before PI becomes PII for an individual known to be in a dataset? Context and rarity play important roles in the answer to this question. The legal tests for PI generally relate to the situation where an individual's identity can "reasonably be ascertained". The definition is very broad and in principle covers any information that relates to an identifiable individual, during their lifetime or for decades after their death [43]. A recent paper published in Nature Communications [44] provides a means to "estimate the likelihood of a specific person to be correctly re-identified, even in a heavily incomplete dataset". The paper is part of a long series showing that only a small number of features need to be linked to identify an individual from a population.

The degree of PI contained in data may be very high (a unique identifier such as a social security number), moderate (surname), low (eye color) or very low (month of birth). It is expected that the PIF in a linked dataset will generally increase as more datasets are linked. Conceptually shown in Figure 9, as more datasets containing PI are linked, a point may be reached where an individual is personally identifiable (a PIF of 1), or "reasonably" identifiable (a PIF within "epsilon" of 1). The "epsilon" is an indication of the difference presented by the gap before the "reasonable" threshold is met.

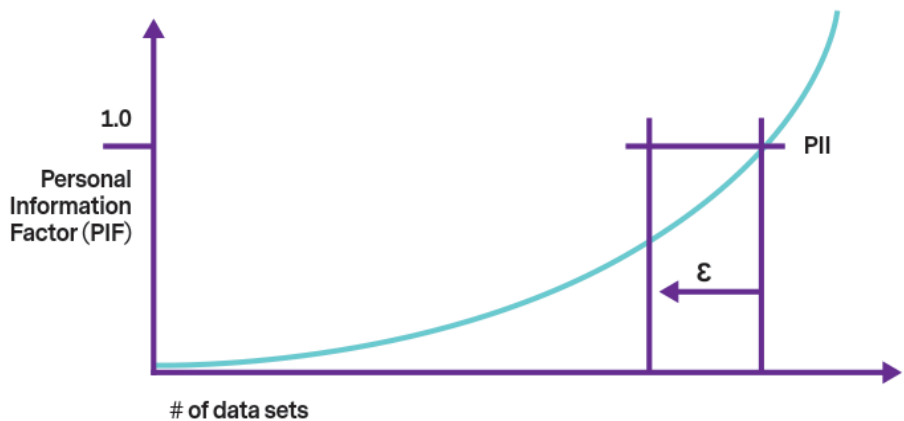

Figure 9: Conceptualization of a PIF and PII threshold point

The focus of this chapter is the call for a quantitative measure and risk framework for "reasonably" in different contexts. In [45], the authors explored a personal information factor (PIF) which is a measure of the PI contained in a linked, deidentified dataset or in the outputs of its analysis. A PIF above a certain threshold (for example, 1.0) means sufficient PI exists to identify an individual: this re-identification risk makes this PII. A value of 0 means there is no PI. It is important to note that the PIF envisaged is not a technique for anonymization; rather, it is a heuristic measure of potential risk of re-identification and of the amount of information which would be revealed from re-identification.

The PIF for both data and the outputs of any analysis based on this data are described in using:

- A measure of the information content of the dataset or the output of the analysis of the data
- The smallest unique group in the dataset or output
- Additional information required to identify an individual from data or output ("epsilon").

There is currently no way to unambiguously determine when linked, deidentified datasets cross the threshold to become personally identifiable. This is a major, unaddressed problem for many digital technologies from AI, and IoT to communications systems as a whole. Applications are relevant in Smart Healthcare, Industrial Automation, and Smart Transportation. Courts in different parts of the world are making decisions about whether privacy is being infringed without formal measures of the level of personal information. This has significant consequences for all future smart services and smart networks.

The relevance specifically for 6G is that, 5G is still largely device / network specific, 6G envisages far more immersive engagement with the network. This means the focus of privacy will move from human-device-network or machine-device-network privacy issues to a much richer set of contextual considerations including time/space/usecase/context. This is an issue which is still to be addressed for 5G. It has largely been ignored to date but GDPR makes it impossible to ignore in future. It is now the subject of ongoing discussion in the standards world.

## 4.3. Privacy-Preserving Technologies

As 5G networks evolve, it is expected that there will be increased reliance on AI enabled smart applications requiring situational, context-aware and customized privacy solutions. Traditional privacy preserving approaches may not be well suited for the future wireless applications due to a diverse and complex set of novel privacy challenges. One potential solution already highlighted in chapter 1 is the use of DLT technologies. DLT technologies such as Blockchain may be an enabler for the use of trustless computing between stakeholders as well as offering privacy protection mechanisms in the network. Blockchain provides security and privacy features such as immutability, transparency, verifiability, anonymity and pseudonymous among other. Blockchain can offer privacy-preserving data sharing mechanisms, optimize the authentication and access control, provide key characteristics such as data integrity, traceability, monitoring, and ensure efficient accountability mechanism among others.

Privacy protection using differential privacy (DP) approaches also seem promising when addressing key challenges that are likely to arise in future intelligent 6G wireless applications. DP operates by perturbing the actual data using artificial design random noise functions before sending the final output to the assigned server [46]. This prevents attackers undertaking a statistical analysis of the received data and prevents inferring personal information from user's data. The concepts related to Federated Learning (FL) are also active topics in the research community for ensuring privacy protection. FL is a distributed machine learning technique that allows model training for large amounts of data locally on its generated source and the required modeling is done at each individual learner in the federation. Instead of sending a raw training dataset, each individual learner transmits their local model to an "aggregator" to build a global model. FL can provide solutions to vital challenges of data privacy, data ownership and data locality as it follows the approach of "bringing the code to the data, instead of the data to the code" [47, 48].

## 4.4. Standardization and Regulatory Aspects

Regulation and standardization work will affect our future technological solutions in 5G networks and beyond. Protection of individual privacy and identity have long been a challenge for standardization bodies since different nations have widely different perspectives and regulation in this area. International standards bodies such JTC 1 have committees working on privacy frameworks and are making progress, but much more remains to be done. Using the European Commission's framework of "Privacy by Design" is one way to frame the challenge. The European Commission implemented decision (20.1.2015) states *"The security industry has thus to face a growing challenge: improving the protection of privacy and personal data, while meeting the requirements of their customers. Whilst legally speaking the customers of the security industry often bear the legal responsibility for complying with data protection rules (being the data controllers), their providers also bear some responsibility for data protection from a societal and ethical point of view. These involve those who design technical specifications and those who actually build or implement applications or operating systems."*

The interconnectedness of global markets means that "Privacy by Design" will likely prevail in many future products and communication techniques, and impact how we will live. Work in standards groups such as JTC 1/WG 11 (Smart cities) reflect this when considering future smart cities [49].

According to the GDPR, if data is collected directly from the data subject, there is a duty to provide the data immediately (Article 13) to the data subject. If the data is collected from a third party (indirectly from the data subject) the information must be provided at the latest after a month from the date of collection (Article 14). If the data subject requests the data, it should be provided upon the request (Article 15). If, as expected, the GDPR continues to provide a data protection framework for 6G, the digital twins of the data subjects may well be assumed to be data subjects as well as their physical, real world counterparts. If the realm of virtual reality was deemed to be a public space, a question arises as to which jurisdiction's laws are applicable. If this question is left without an answer, the choice may well be determined by the private sector. Big tech companies already have significant data and control power and are well prepared to decried how the digital world will operate through control of data and even through digital currencies. If this question is left unanswered; human rights may be jeopardized and 6G may not achieve the expected social benefits.

A future which sees us working via our digital avatars or by telepresence will also have health and safety issues. While the right of employees to 'disconnect' is beginning to be accepted as a legal norm through some European countries such as France, the EU is preparing to formalize this [50]. Finland's Working Hour Act came into force January 2020. This Act allows workers the right to determine how and where to work for half of their yearly working time. 6G communication is expected to have benefits for workers on this flexible working system. However, new legal consequences may arise around industrial relations matters in a digital, mirror workplace where telepresence and avatars are essential parts of the production system where human employees work in collaboration. Furthermore, the Future digital and physical worlds will be deeply entangled, malicious cyber activities could lead to loss of property and life. Attribution of responsibility in case of a physical harm caused by a digital twin or an automated system against a human is a compelling legal issue. A critical question will always remain: Who is liable? After answering this question the next is the type of liability which will be applied to the case. Liability occurs either from a contractual relationship or tort or unjust enrichment. Tort on the other hand, may depend on fault (by intention or by negligence) or strict liability. Multiple liability issues may appear related to Work health and safety issues (physical harm to life of human or to workplace), data protection and cyber security.

## 4.5. Research challenges for privacy in 6G

- Develop a measure of Personal Information (PI) in linked identified data sets used to provide smart services
- Develop a threshold test for Personally Identifiable Information (PII)
- Consider the impact of temporal, spatial, contextual settings in the measures of PI
- What can 5G privacy protection approaches offer to 6G systems?
- Develop frameworks to understand the trade-offs between privacy and trust in 6G systems

**Chapter 4: Privacy protection in 6G: principles, technologies and regulation**